\documentclass{article}

\usepackage{graphicx}

\begin{document}

\title{Analytical Treatment of Benzene Transmittivity}

\date{July 20, 2016}

\author{Kenneth W. Sulston\thanks{email: sulston@upei.ca} 
\thanks{School of Mathematical and Computational Sciences, University of Prince Edward Island, Charlottetown, PE, C1A 4P3, Canada}\and Sydney G. Davison\thanks{email: sgdaviso@uwaterloo.ca} \thanks{Department of Applied Mathematics, University of Waterloo, Waterloo, ON, N2L 3G1, Canada} \thanks{Department of Physics and the Guelph-Waterloo Physics Institute, University of Waterloo Campus, Waterloo, ON, N2L 3G1, Canada}}

\maketitle

\section{Abstract}

The renormalization method has been previously used in conjunction with the Lippman-Schwinger
equation to calculate transmission probabilities for benzene molecules, within the
tight-binding approximation.
Those results are extended here by presenting explicit formulas for the transmission
probability functions, which are subsequently analysed for their anti-resonances,
resonances and other maxima and minima.

\section{Introduction}

The goal of nanoelectronics is to construct circuits at the molecular/atomic level \cite{ref1}.
To this end, one of the most important molecules is benzene, because of its structural 
simplicity and its appearance in larger compounds.
Consequently, a detailed understanding of its electron-transmission properties is central
to delving further into those of more complicated structures.
In particular, when feasible, analytical solutions can be of considerable value.
Of many studies of benzene, one recent one is that of Hansen {\it et al} \cite{ref2},
who used the L\"{o}wdin partitioning technique to derive the Green's function, in analytical
form, for an isolated benzene molecule.
Subsequently, the coupling of the molecule to the metallic leads was described using first-order
 perturbation theory.
This allowed the transmission function $T(E)$ to be calculated, with particular attention
paid to the anti-resonances (for which $T=0$).
Even more recently, Dias and Peres \cite{ref3} used the Green's function method, within
the tight-binding approximation, to derive an analytical expression for the transmission function
through a para-benzene ring, albeit in a non-simplified form.

\begin{figure}[h]
\includegraphics[width=12cm]{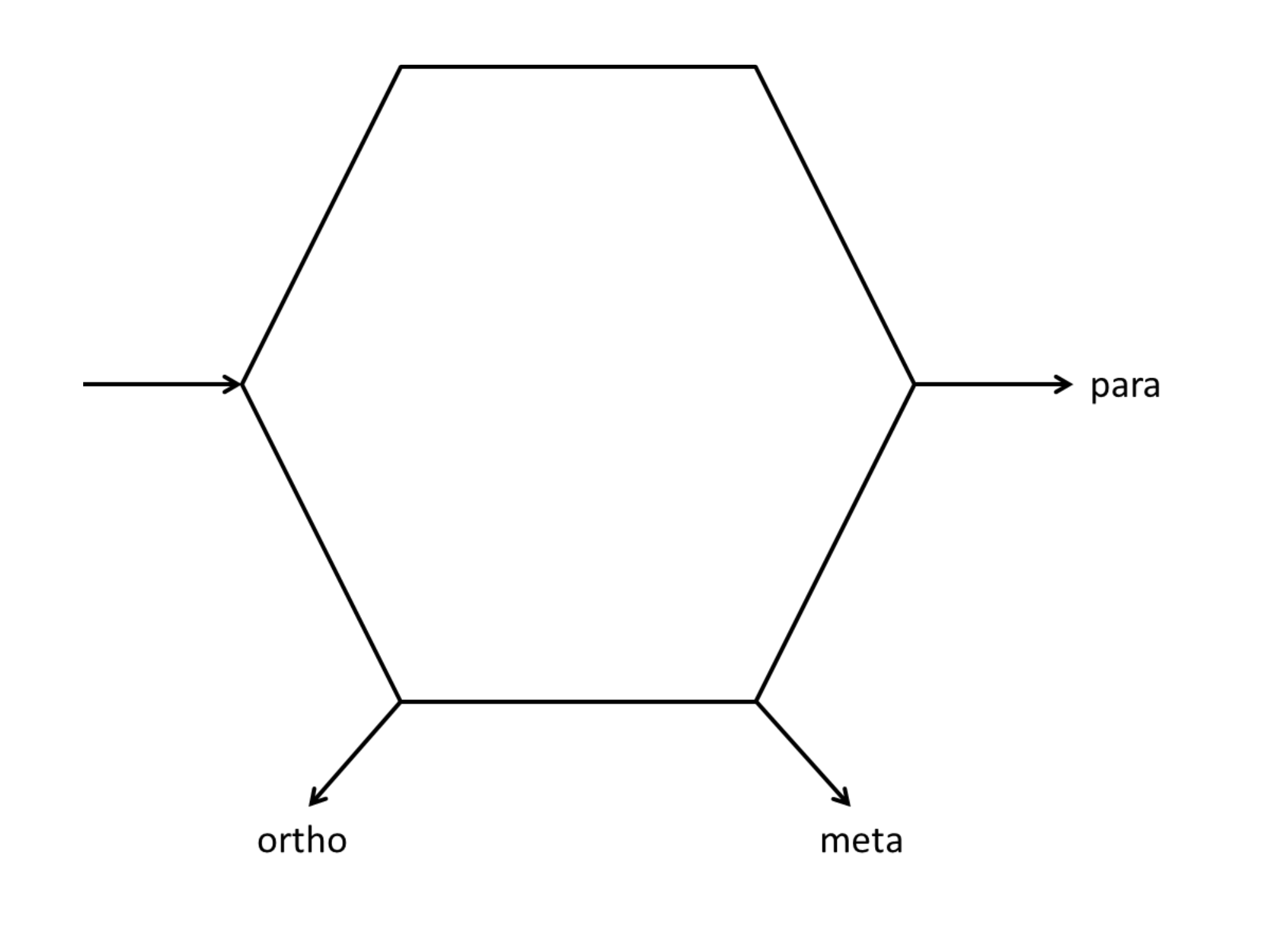}
\caption{Benzene molecule showing para, meta and ortho connections to leads.}
\label{fig1}
\end{figure}

Our own studies of benzene systems  are based on applying the renormalization method \cite{ref4}
to the tight-binding Green's function formalism, which although equivalent to that of \cite{ref3}, has
the advantage of circumventing the direct evaluation of certain integrals. 
As in \cite{ref3}, the transmission function is then evaluated via the Lippmann-Schwinger equation \cite{ref5}.
In addition to investigating transmission through a single benzene molecule \cite{ref5}, the method
has also been used to study series and parallel circuits of such molecules \cite{ref5}, overlap
effects \cite{ref6}, replacement of a carbon atom \cite{ref7}, and applied-field effects \cite{ref8}.
In this paper, which can be viewed as a supplement to \cite{ref5}, we present the analytical forms
of the transmission functions for a single benzene molecule, connected  to carbon leads, in each
of the para, meta and ortho configurations (see Figure \ref{fig1}).
Additionally, particular attention is paid to the key analytical features of each function, namely,
anti-resonances (which are the zeroes of $T$), resonances (for which $T=1$), and other minimum 
and maximum values.
These details can be of considerable value in analysing more complicated systems, based
on benzene, but for which explicit analytical solutions are not feasible.

\section{Summary of Previous Work}

Our starting point is a brief summary of the relevant results from \cite{ref5} for transmission
through a single benzene molecule;
for complete details, we refer the reader to \cite{ref5}.
The basic method therein was to use the Lippmann-Schwinger equation to derive the 
transmission probability for an electron through a one-dimensional tight-binding chain,
containing a double impurity. 
Subsequently, this allows benzene (and indeed, a wide variety of systems) to be studied,
by using the renormalization technique to reduce the molecule to a dimer, with rescaled
energy-dependent parameters, which then plays the role of the double impurity.
The main mathematical result is that the transmission-energy probability function has the
form
\begin{equation}
T(X(E))= {{(1+2\gamma)^2 (4-X^2)} \over {(1-2Q)^2 (4-X^2) + 4(P-QX)^2}}  ,
\label{eq1}
\end{equation}
where
\begin{equation}
P = z_0+z_1  ~,~ Q = z_0 z_1 - \gamma -\gamma^2 ,
\label{eq2}
\end{equation}
with
\begin{equation}
z_{0,1} = ({\alpha}_{0,1} - \alpha)/ 2\beta ~,~ \gamma = ({\beta}_{01} -\beta)/ 2\beta ,
\label{eq3}
\end{equation}
and the reduced dimensionless energy is
\begin{equation}
X= (E - \alpha)/\beta .
\label{eq4}
\end{equation}
In the above, $\alpha$ and $\beta$ are the site and bond energies, respectively, for
the carbon atoms in benzene and in the leads, while $\alpha_0$, $\alpha_1$ and
$\beta_{01}$ are the rescaled parameters for the dimer.
In general situations, the dimer is asymmetric resulting in $\alpha_0 \ne \alpha_1$, but
for the cases  considered here, it turns out that the dimers are all symmetric, so that
$\alpha_0 = \alpha_1$, and hence $z_0=z_1$ in (\ref{eq3}).

\section{Analysis of $T(E)$ curves}

\subsection{Para-benzene}

\begin{figure}[h]
\includegraphics[width=12cm]{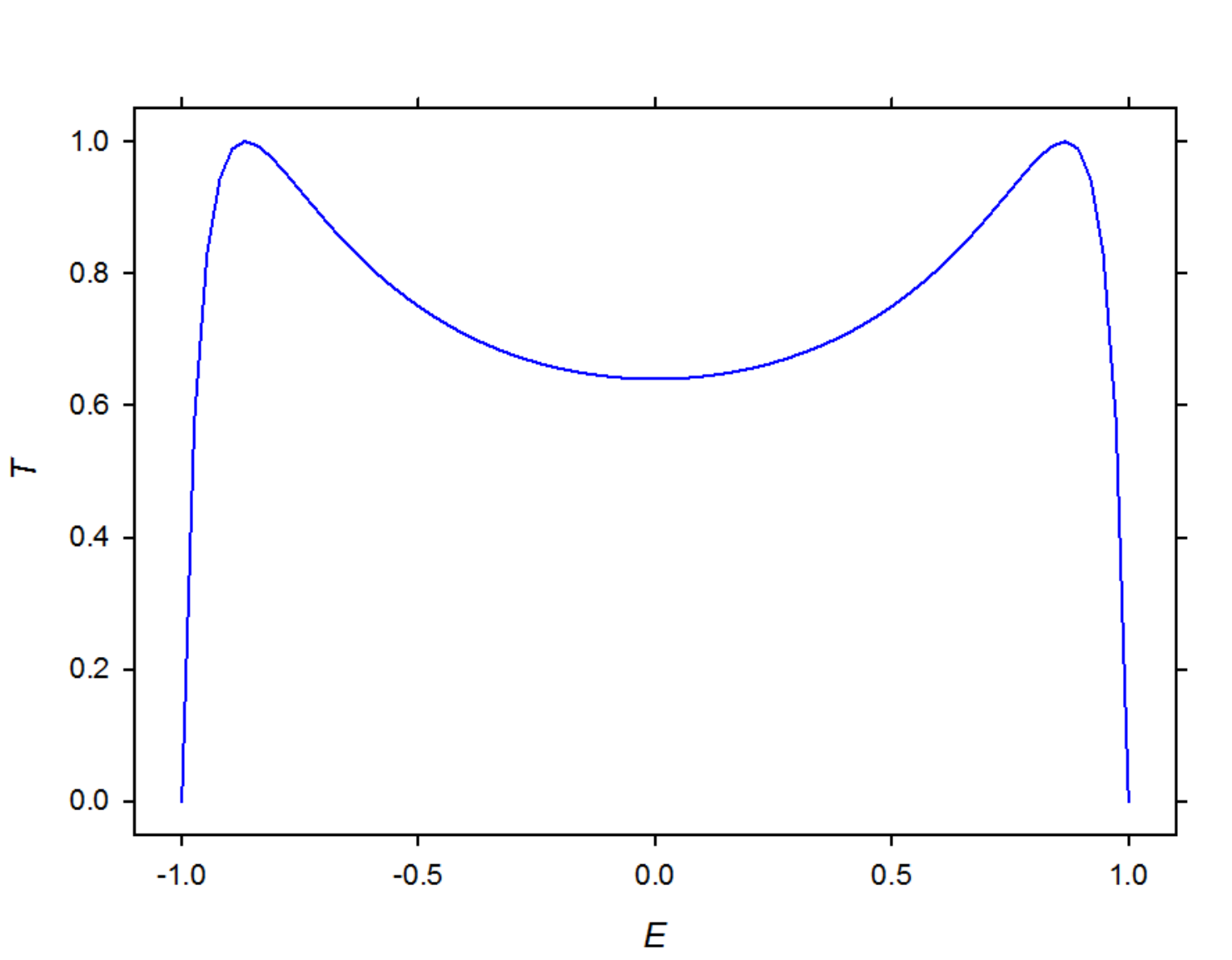}
\caption{Transmission probability $T$ versus energy $E$ for para-benzene.}
\label{fig2}
\end{figure}

As derived in detail in equations (29) and (30) of \cite{ref5}, the rescaled dimer parameters
for para-benzene are
\begin{equation}
\alpha_0 \equiv \alpha_1=  \alpha + \beta_{01} X ,
\label{eq5}
\end{equation}
\begin{equation}
\beta_{01} = 2 \beta (X^2-1)^{-1} ,
\label{eq6}
\end{equation}
For convenience of presentation, we use parameter values of $\alpha=0$ and
$\beta=-1/2$, for which the energy band is the range $-1 \le E \le 1$.
On substituting  (\ref{eq5}) and  (\ref{eq6}) into  (\ref{eq1}), and using  (\ref{eq2}) to (\ref{eq4}),
some lengthy algebraic manipulation leads to the comparatively simple expression for the
transmission probability
\begin{equation}
T(E) = {{16 (1-E^2)} \over {(5-4E^2)^2}},
\label{eq7}
\end{equation}
whose graph is shown in Figure \ref{fig2}, and is clearly seen to be symmetric, in
accord with $T(E)$ being an even function.

\vspace{8pt}
\noindent (a) {\bf Anti-resonances}

Anti-resonances are energies for which the transmission probability is 0, so setting
$T(E)=0$ in (\ref{eq7}) yields only $E= \pm 1$, and correspond to the band
edges, which are not considered to be true anti-resonances. 
This is in accord with what is seen in Figure \ref{fig2}.

\vspace{8pt}
\noindent (b) {\bf Resonances}

Resonances are energies for which  the transmission probability is 1, which from 
(\ref{eq7}) requires that 
\begin{equation}
16 (1-E^2) = (5-4E^2)^2,
\label{eq8}
\end{equation}
which can be rearranged as 
\begin{equation}
(4E^2-3)^2=0,
\label{eq9}
\end{equation}
from which it is clear that the only resonances occur at 
\begin{equation}
E = \pm \sqrt 3 /2 = \pm 0.866.
\label{eq10}
\end{equation}
This pair of resonances is clearly visible on Figure \ref{fig2}.

\vspace{8pt}
\noindent (c) {\bf Extrema}

Extreme values of $T(E)$ include resonances and anti-resonances, as well as other
maxima and minima.
These are determined by setting $T'(E)=0$, which produces
\begin{equation}
-32E (4E^2-3) (5-4E^2)^{-3} = 0,
\label{eq11}
\end{equation}
whose solutions are $E = \pm \sqrt 3 /2$ and $E=0$.
The first two critical points are obviously the resonances (maxima) mentioned above,
while $E=0$ corresponds to a minimum value of $T(0)=16/25$, as can be seen in
Figure \ref{fig2}, and can be verified using the second-derivative test.

\vspace{8pt}
Thus we can confirm analytically all the key features of $T(E)$ seen on its graph.
It is important (and reassuring) to note that Figure \ref{fig2} matches precisely the
corresponding Figure 6(a) of Dias and Peres \cite{ref3},  which was produced using
the same model and parameter values, but with somewhat different methodology.
Moreover, there is also agreement with the work of Hansen {\it et al} \cite{ref2},
despite some differences in the models used.
Specifically, Figure 4 of that paper indicates that the $T(E)$ curve for para-benzene
is without anti-resonances.

\subsection{Meta-benzene}

\begin{figure}[h]
\includegraphics[width=12cm]{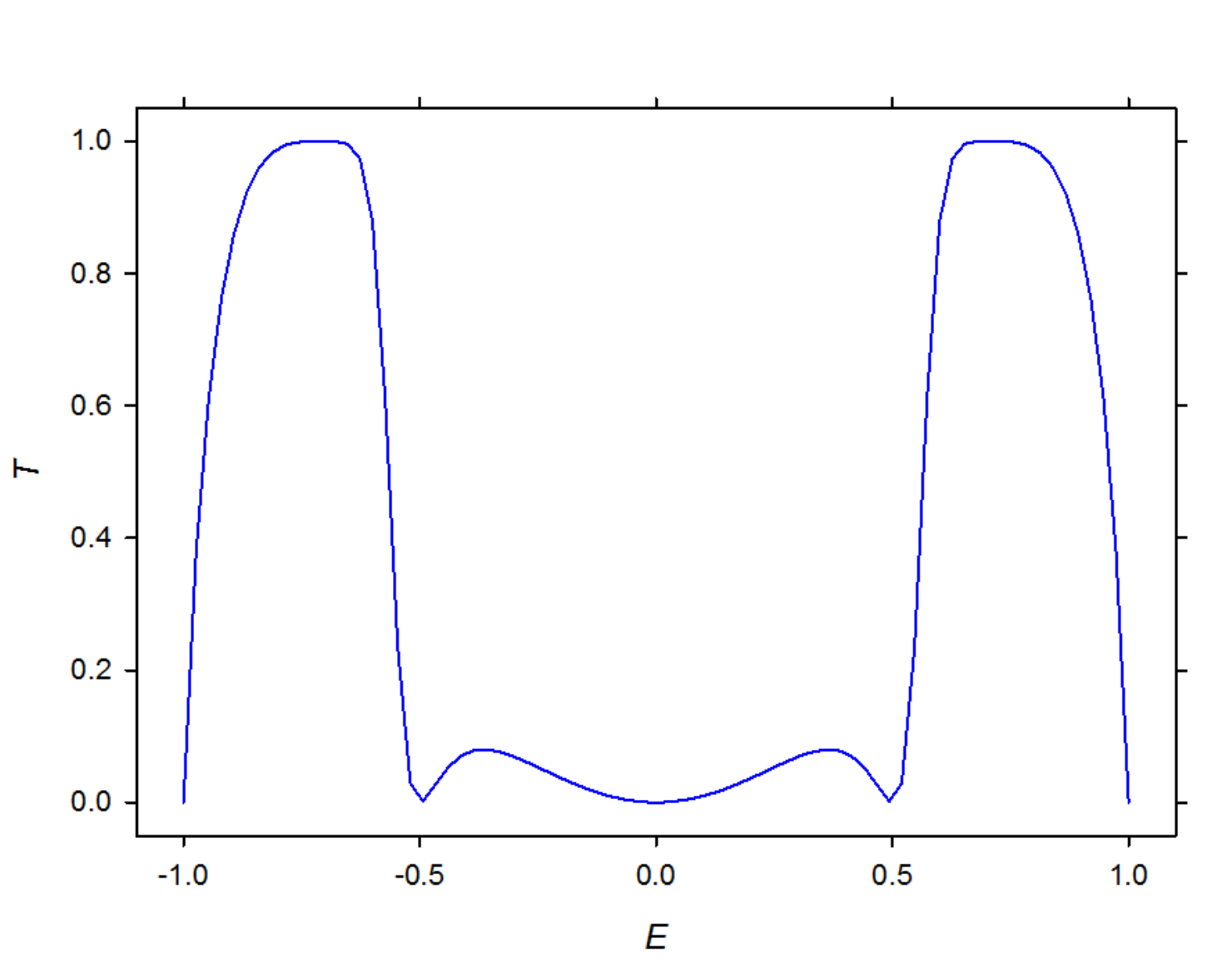}
\caption{Transmission probability $T$ versus energy $E$ for meta-benzene.}
\label{fig3}
\end{figure}

Once again referring to \cite{ref5} (equations (32) and (33)), the rescaled parameters
for meta-benzene are
\begin{equation}
\alpha_0 \equiv \alpha_1 = \alpha + \beta X^{-1} + \beta_{01}  ,
\label{eq12}
\end{equation}
\begin{equation}
 \beta_{01} =  \beta X^{-1} (X^2-1) (X^2-2)^{-1}.
\label{eq13}
\end{equation}
Substituting (\ref{eq12}) and (\ref{eq13}) into (\ref{eq1}), and using Maple to
perform the simplifications, leads to
\begin{equation}
T(E) = {{E^2 (1-E^2)(1-4E^2)^2} \over {1+E^2(1-E^2)(8E^2-7)}},
\label{eq14}
\end{equation}
which is graphed in Figure \ref{fig3}.
Although more complicated than that for para-benzene, the graph for meta-benzene
is once again symmetric.

\vspace{8pt}
\noindent (a) {\bf Anti-resonances}

Putting $T(E)=0$ in (\ref{eq14}) sets its numerator to 0, from which we obtain the
band edges $E=\pm 1$, as well as 3 anti-resonances
\begin{equation}
E =0 ~,~E = \pm 1 /2,
\label{eq15}
\end{equation}
which are clearly visible in Figure \ref{fig3}.
We note that these anti-resonances correspond precisely to those found for meta-benzene
by Hansen {\it et al} \cite{ref2}.

\vspace{8pt}
\noindent (b) {\bf Resonances}

To find resonances, we set $T(E)=1$ in (\ref{eq14}), i.e., set the numerator equal
to the denominator, which produces a degree-8 equation which surprisingly simplifies to just
\begin{equation}
(2E^2-1)^4=0.
\label{eq16}
\end{equation}
The only solutions of (\ref{eq16}) are the two resonances
\begin{equation}
E = \pm \sqrt 2/2 = \pm 0.707,
\label{eq17}
\end{equation}
which can be seen in Figure \ref{fig3}.

\vspace{8pt}
\noindent (c) {\bf Extrema}

As usual, critical points are found by setting $T'(E)=0$, which in application to (\ref{eq14})
leads, after much computer algebra, to the condition
\begin{equation}
2E(2E+1)(2E-1)(2E^2-1)^3(2E^2+2E-1)(2E^2-2E+1)=0.
\label{eq18}
\end{equation}
Equation (\ref{eq18}) has a total of 9 distinct real solutions.
Easily seen are the anti-resonances $E =0$ and $E = \pm 1 /2$, from (\ref{eq15}), which are obviously
minima.
Also readily apparent are the two resonances, $E = \pm \sqrt 2/2$, encountered in  (\ref{eq17}).
Lastly, the pair of quadratic factors at the right end of (\ref{eq18}) admit 4 real roots, only 2 of
which lie in the transmission band, namely
\begin{equation}
E = \pm (\sqrt 3 -1)/2  = \pm 0.366.
\label{eq19}
\end{equation}
The second-derivative test can confirm that the solutions (\ref{eq19}) are maxima, although
not resonances, as they can be clearly seen on the graph of Figure \ref{fig3} as the energies
corresponding to the smaller inner pair of peaks.

\vspace{8pt}
In summary, we have been able to verify analytically all the major features of Figure \ref{fig3},
and to reaffirm the relevant part of the work of Hansen {\it et al} \cite{ref2}.
(Unfortunately, Dias and Peres \cite{ref3} only considered para-benzene.)

\subsection{Ortho-benzene}

\begin{figure}[h]
\includegraphics[width=12cm]{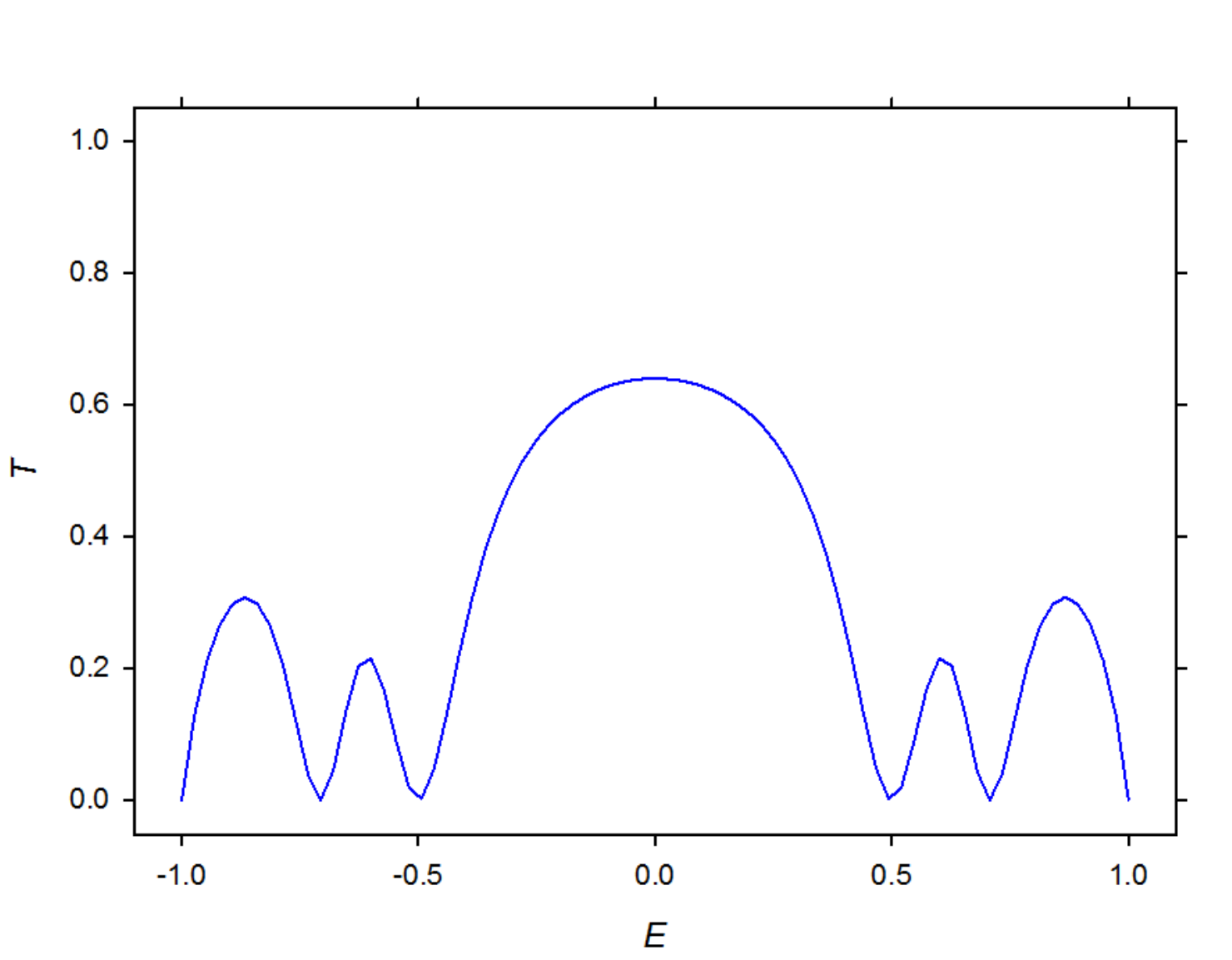}
\caption{Transmission probability $T$ versus energy $E$ for ortho-benzene.}
\label{fig4}
\end{figure}

Lastly, we turn to ortho-benzene, for which the rescaled parameters are (equations (34) and (35) of \cite{ref5})
\begin{equation}
\alpha_0 \equiv \alpha_1=  \alpha + \beta (X^2-2) (X^2-X-1)^{-1} - \beta_{01}  ,
\label{eq20}
\end{equation}
\begin{equation}
 \beta_{01} =  \beta (X^2-1) (X^2-2) [(X^2-1)^2-X^2]^{-1}.
\label{eq21}
\end{equation}
Following our earlier strategy, we substitute (\ref{eq20}) and (\ref{eq21}) into (\ref{eq1}),
then use Maple to produce the transmission probability function
\begin{equation}
T(E) = {{16 (4E^2-1)^2 (2E^2-1)^2 (E^2-1)} \over {[4(2E+1)(2E-1)^2(E+1)^2+1][4(2E-1)(2E+1)^2(E-1)^2-1]}},
\label{eq22}
\end{equation}
which is obviously a more complicated expression than those for  the para and meta cases,
and results in the more elaborate graph shown in Figure \ref{fig4}.
In accord with the two previous cases, the graph here is again symmetric.

\vspace{8pt}
\noindent (a) {\bf Anti-resonances}

The form of $T(E)$ given in  (\ref{eq22}) easily yields its zeroes, which are the band edges
$E= \pm 1$ as well as 2 pairs of anti-resonances, specifically,
\begin{equation}
E = \pm 1/2 ~,~ E = \pm \sqrt 2/2 = \pm 0.707,
\label{eq23}
\end{equation}
which can be seen in Figure \ref{fig4}.
Once again, these results agree with the work of Hansen {\it et al} \cite{ref2}.

\vspace{8pt}
\noindent (b) {\bf Resonances}

Setting $T(E)=1$ leads to an 8th-degree equation, which can be
shown to factorize as
\begin{equation}
(16 E^4-12E^2+3)^2=0.
\label{eq23a}
\end{equation}
Hence (\ref{eq23a}) can be solved to show that it has no real solutions, so there are no resonances.
This is in agreement with the graph.

\vspace{8pt}
\noindent (c) {\bf Extrema}

Starting from $T(E)$ in (\ref{eq22}) to set up the equation $T'(E)=0$ leads, after
some computer algebra, to the factorized critical-point condition
\begin{equation}
2E(2E+1)(2E-1)(2E^2-1)(4E^2-3)(16E^4-12E^2+3)(32E^6-24E^4+18E^2-5)=0,
\label{eq24}
\end{equation}
which certainly allows for a lot of solutions, not all of them real.
First, the 4 anti-resonances (\ref{eq23}) can be recovered as the zeroes of the linear factors
$2E \pm 1$ and the lead quadratic factor $2E^2-1$.
Next, the leading linear factor $E$ obviously yields a critical point at
\begin{equation}
E=0,
\label{eq25}
\end{equation}
which can be seen on Figure \ref{fig4} to correspond to the large maximum at the 
center of the graph.
The second quadratic factor, $4E^2-3$, gives rise to a pair of solutions, namely,
\begin{equation}
E= \pm \sqrt 3 /2 = \pm 0.866,
\label{eq26}
\end{equation}
which can be seen on the graph to be the outer pair of maxima.
The quartic factor has no real zeroes.
Finally, the sextic factor leads to only a pair of real  zeroes, which are 
\begin{equation}
E= \pm [(1/8)(12+4\sqrt{41})^{1/3}-(12+4\sqrt{41})^{-1/3}+1/4]^{1/2}  = \pm 0.609,
\label{eq27}
\end{equation}
corresponding on the graph to the smallest (and inner) pair of maxima.
The set of 5 energies given by (\ref{eq25})-(\ref{eq27}) completes the group of maxima
seen in Figure \ref{fig4}, with none of them being resonances, as was pointed out earlier.

\vspace{8pt}

Despite the complicated form of the $T(E)$ function of (\ref{eq22}), we have,
with the assistance of computer algebra, been able to extract all of its key features,
seen in Figure \ref{fig4}.

\section{Conclusions}

In this paper, we have followed up our previous work on transmission through benzene
molecules, by deriving explicit analytic formulas for the transmission probability functions,
for each of the three types of lead-connections. 
Subsequently, this allowed us to determine analytically all the key features seen on the
graphs of these functions, specifically, anti-resonances, resonances and other extrema.
Where possible, the work here was compared with that of other researchers and found
to be in agreement.

\section{Keywords}

benzene, electron transmission, Green's functions, renormalization method, analytical solutions

\end{document}